\documentclass[aps,eqsecnum,pra]{revtex4}
\usepackage{graphics,graphicx}
\usepackage{amsmath}
\usepackage{amssymb,latexsym,mathrsfs}
\usepackage{hyperref}

\def\bea{\begin{eqnarray}}
\def\eea{\end{eqnarray}}
\def\ba{\begin{array}}
\def\ea{\end{array}}

\def\beq{\begin{equation}}
\def\eeq{\end{equation}}

\begin{document}

\title{Laser trapping of ions and asymptotic minimization of Decoherence}

\author{Samyadeb Bhattacharya$^{1}$ \footnote{sbh.phys@gmail.com}, Sisir Roy$^{2} $ \footnote{sisir@isical.ac.in}}
\affiliation{$^{1,2}$Physics and Applied Mathematics Unit, Indian Statistical Institute, 203 B.T. Road, Kolkata 700 108, India \\}

\vspace{2cm}
\begin{abstract}

\vspace{1cm}

Decoherence time has been calculated for an optical ion trap of Be atoms in a bistable potential model. Comparison has been made between decoherence time and Zeno time for double well potential as a special case. Zeno time is considered as a lower limit of decoherence time for sustainable quantum coherence. Equality of the respective timescales provides a certain transitional temperature, below which decoherence can be asymptotically minimized.
\vspace{2cm}


\vspace{1cm}

\end{abstract}

\vspace{1cm}

\maketitle

\section{Introduction}

Quantum memory can store information in superposition states of a collection of two-level systems. Optical ion trap by laser cooling has been prepared to construct quantum logic gates \cite{5,6}. In those systems, negative role played by quantum decoherence \cite{1} is quite significant. Randomization of the quantum states produced by entanglement with environmental modes is inevitable in case of storage or processing of non-orthogonal states and environmental interaction allows leakage of some information to the environment \cite{1a}. Since it is practically impossible to disentangle the system from the environment, our main efforts are focussed on minimizing decoherence. In this attempt of decoherence minimization, Zeno dynamics plays a very significant role \cite{1b,1c}. Quantum Zeno effect \cite{3,4} is depicted as the complete freezing of the decay dynamics due to frequent measurement. It has been shown previously that very frequent measurement of excited states can suppress the decoherence \cite{7}. In our understanding decoherence and Zeno effect has got intrinsic reciprocal relationship between them. The argument behind this statement is as follows: whenever any disturbance in the form of measurement dominates the time evolution of the state of the system, the system is forced to evolve in a subspace of the total Hilbert space \cite{7a}. This subspace is called ``Zeno subspace". Nonselective measurement causes the appearance of these subspaces. Facchi et.al \cite{7a} have shown that frequent nonselective measurement splits the total Hilbert space into invariant quantum zeno subspaces, between which probability leakage is not possible. But probability is conserved within each subspace. So each of the subspace can be considered as an reduced isolated system. If the system undergoes very strong environmental interaction, due to extreme decoherence, these isolated subspaces may not be sustainable. So we can infer that the zeno effect characterized by a certain time scale (Zeno time), gives a kind of lower limit to decoherence, below which the process of decoherence will be uncontrollable. The relation between these two phenomena is reciprocal in the sense that within the zeno subspace, due to it's isolated nature, it precludes environment induced decoherence. Exploiting this relation, we will formulate the procedure to compare the respective time scales and come up with a certain transitional temperature, below which asymptotic minimization of state decoherence is possible.  \\
The master equation for the density operator in position representation of a certain quantum system can be given as \cite{1}
\beq\label{1.1}
\dot{\rho}= -\frac{i}{\hbar}\left[H,\rho\right]-\gamma \left(\frac{\partial}{\partial x}-\frac{\partial}{\partial x'}\right)-\frac{2m\gamma K_B T}{\hbar^2}(x-x')^2 \rho
\eeq
where the first term on the right hand side is the usual commutator term of the von Neumann equation. The second term represents dissipation with $\gamma$ as the relaxation rate. The third and last term represent the fluctuations leading to random Brownian effects. This term being proportional to $(x-x')^2$, though has little effect on the diagonal peaks of the density matrix, but affects the off-diagonal peaks considerably and causes them to decay. Hence the effect of this last term leads to the destruction of quantum coherence. From equation (\ref{1.1}) we can easily get that the decay rate of the off-diagonal peaks of the density matrix
\beq\label{1.2}
\frac{d\rho}{dt}=-\frac{2m\gamma K_B T}{\hbar^2}(x-x')^2 \rho = -\tau_{dec} ^{-1}\rho
\eeq
where
\beq\label{1.3}
\tau_{dec}=\frac{\hbar^2}{2m\gamma K_B T (x-x')^2}
\eeq
is the time scale on which the quantum coherence disappears and is defined as decoherence time. From the solution of equation (\ref{1.2}), one can easily get
\beq\label{1.4}
\rho(x,x',t)=\rho(x,x',0) \exp(-t/\tau_{dec})
\eeq
Decoherence visibly supresses the interference between macroscopically different quantum states, which is precisely the very property that distinguishes quantum mechanics from it's classical counterpart from observational perspective. Here we will consider tunneling in a bistable potential as a model system to develop the expression for decoherence time. As a physically realistic example we will consider a system of laser cooled trapped ion \cite{5}, where decoherence appears in the dynamics of hyperfine states. Comparison between decoherence time and zeno time for this specific case will lead us to find  the transitional temperature over which decoherence will dominate the whole process.\\

\section{Decoherence-Dwell time ratio: Condition to control decoherence}

Let us first concentrate on the calculation of the relaxation rate $(\gamma)$ in presence of dissipative interaction. In a recent paper \cite{8} we have estimated the weak value of dwell time for a dissipative spin-half system using the same formalism. The approach that has been used here, was originally developed by Caldirola and Montaldi \cite{2} introducing a discrete time parameter ($\delta$) incorporating the properties of environment. The Schr\"{o}dinger difference equation in presence of environment induced dissipation is given by
\beq\label{2.01}
H_i|\psi\rangle=i\hbar\frac{[|\psi(t)\rangle-|\psi(t-\delta)\rangle]}{\delta}
\eeq
It has been shown \cite{2} that this equation has retarded nature and so naturally implies the dissipative character of it's solution. The discrete time parameter ($\delta$) appears as some sort of relaxation time, incorporating the environment induced dissipation. To supplement this difference equation, we will show further that the time parameter ($\delta$) can be expressed as a function of the energy eigen-values of the quantum states. Now as a consequence of the retarded nature of eqn (\ref{2.01}), we can see that the ground state will also decay. So to stabilize the ground state, the Schr\"{o}dinger difference equation is scaled as
\beq\label{2.1}
(H_i-H_0)|\psi\rangle=i\hbar \frac{|\psi(t)\rangle-|\psi(t-\delta)\rangle}{\delta}
\eeq
where $H_i$ and $H_0$ are the Hamiltonian for i-th and ground state respectively. $H_0$ is introduced in the equation to stabilize the ground state \cite{2}.
We expand $|\psi(t-\delta)\rangle$ in Taylor series to get
\beq\label{2.2}
(H_i-H_0)|\psi\rangle=i\hbar \frac{[1-e^{\delta\frac{\partial}{\partial t}}]|\psi(t)\rangle}{\delta}
\eeq
Setting the trial solution as $|\psi(t)\rangle=e^{-\alpha t}|\psi(0)\rangle$ and solving for $\alpha$, we get
\beq\label{2.3}
\alpha= \frac{1}{\delta}\ln\left(1+i(E_i-E_0)\delta/\hbar\right)
\eeq
where $E_i$ and $E_0$ are the eigenvalues for the corresponding hamiltonians. Expanding the logarithm upto third order, we find that the time evolution takes the form
\beq\label{2.4}
\exp\left[-i\left(\frac{(E_i-E_0)}{\hbar}-\frac{(E_i-E_0)^3\delta^2 }{\hbar^3}\right)t-\frac{(E_i-E_0)^2\delta }{\hbar^2}t\right]
\eeq
So from (\ref{2.4}) we find the decay rate as
\beq\label{2.5}
\gamma=\frac{(E_i-E_0)^2\delta}{\hbar^2}
\eeq
Setting the final Hamiltonian as $H_f$, we also find the unknown time parameter as \cite{8}
\beq\label{2.6}
\delta=\frac{\hbar}{E_i-E_0}\sqrt{\frac{E_i-E_f}{E_i-E_0}}
\eeq
So using the value of the time parameter $\delta$ in equation (\ref{2.5}), we find the decay constant
\beq\label{2.7}
\gamma=\frac{\sqrt{(E_i-E_f)(E_i-E_0)}}{\hbar}
\eeq
This is the decay constant for a quantum system decaying from initial to final energy eigenstate denoted by $H_i$ and $H_f$ respectively. We will substitute this decay (relaxation) constant in the expression of decoherence time given by equation (\ref{1.3}). The interaction parameter was given by $\delta$, which was substituted by a function of the initial and final Hamiltonians considering the time evolution dynamics. But to calculate the spatial shift ($\Delta x= x-x'$), we have to consider the spatial dynamics of the system. Here we will focus our attention on an asymmetric double well potential approximated as a two-state system with considering only the ground states of the wells separated by an asymmetry energy $\epsilon$. We construct our model on the demonstration of a quantum logic gate prepared by a trapped ion laser cooled to the zero point energy \cite{5}. In this particular case, the target qubit is spanned by two $^{2}S_{1/2}$ hyperfine ground states ($|\uparrow\rangle~ \mbox{and}~|\downarrow\rangle$ states) of a single $^{9}Be^{+}$ ion separated by $\nu_0=1.250$ GHz. We set these two energy levels as the ground states of the two wells of the double well potential separated by the asymmetry energy $\epsilon=h \nu_0$. The control qubit $|n\rangle$ is spanned by the first two states of trapped ion ($|0\rangle~ \mbox{and}~ |1\rangle$), which can be identified by the first two states of each well approximated as harmonic oscillators. These two states are separated by $\nu_x\simeq11$ MHz.So the basic four eigenstates are given by $|n\rangle|S\rangle=|0\rangle|\uparrow\rangle,|0\rangle|\downarrow\rangle,|1\rangle|\uparrow\rangle,|1\rangle|\downarrow\rangle$.
\newline Let us now consider a quartic potential of the form \cite{9}
\beq\label{2.8}
V(x)=\frac{1}{2}m\omega^2 x^2 \left[\left(\frac{x}{a}\right)^2-A\left(\frac{x}{a}\right) +B\right]
\eeq
where $A$ and $B$ are dimensionless constants. For the particular case of double well, $A=14$ and $B=45$.
\begin{figure}[htb]
{\centerline{\includegraphics[width=7cm, height=5cm] {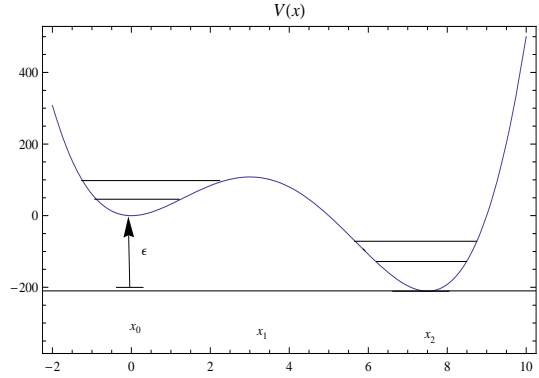}}}
\caption{V(x) vs. x with parameters $A=14$ and $B=45$}
\label{figVr}
\end{figure}

The two potential minima can be found at
\beq\label{2.9}
x_0=0~~~~~~~\mbox{and}~~~~~~~x_2=\frac{a}{8}\left[3A+\sqrt{9A^2-32B}\right]
\eeq
These two minima are separated by the barrier maxima situated at
\beq\label{2.10}
x_1=\frac{a}{8}\left[3A-\sqrt{9A^2-32B}\right]
\eeq
The potential can be expressed in terms of dimensionless variable $\xi=x/a$ as
\beq\label{2.11}
V(\xi)=\frac{1}{2}m\omega^2a^2 \xi^2[\xi^2-A\xi+B]
\eeq
Now the dimensionless form of the Hamiltonian is given by
\beq\label{2.12}
K=\frac{\beta^2}{2}\left[p^2+\xi^2(\xi^2-A\xi+B)\right]
\eeq
where $H=\hbar \omega K$ is the actual Hamiltonian and the parameter $\beta^2=\frac{m\omega a^2}{\hbar}$. Here, the wells are approximated as harmonic oscillators. So we expand the potential given by equation (\ref{2.11}) around the first minima $\xi=0$ to find
\beq\label{2.13}
V({\xi})=\frac{1}{2}\beta^2B\xi^2
\eeq
The normalized ground state wave function for this approximated potential can be set as
\beq\label{2.14}
\psi(\xi)=\left(\frac{\nu}{\pi}\right)^{1/2}\exp\left[-\frac{1}{2}\nu\xi^2\right]
\eeq
where $\nu=\sqrt{B}\beta^2$.\\
\noindent Let us consider the transition from left well to the right well. So here the initial position of the particle is $x'=0$. So $\Delta x=x$. Now
\beq\label{2.15}
\begin{array}{ll}
<x^2>=a^2<\xi^2>=\sqrt{\frac{\nu}{\pi}} a^2 \int_{-\infty}^{\infty} \xi^2e^{-\nu \xi^2}d\xi \\
~~~~~~~~ = \frac{a^2}{2\nu}
\end{array}
\eeq
Therefore the average of the square of the spatial shift
\beq\label{2.16}
\Delta x^2=<x^2>=\frac{a^2}{2\nu}
\eeq
By using (\ref{2.7}) and (\ref{2.16}) in equation (\ref{1.3}), we deduce the decoherence time as
\beq\label{2.17}
\tau_{dec}=\frac{\hbar \sqrt{B}}{\sqrt{(E_i-E_f)(E_i-E_0)}}\left(\frac{\hbar\omega}{K_B T}\right)
\eeq
Now the asymmetry energy ($\epsilon$) can be expressed as
\beq\label{2.18}
\epsilon=V(x_0)-V(x_2)
\eeq
For a double well potential ($A=14 ~\mbox{and}~ B=45$), from equation (\ref{2.18})we get
\beq\label{2.19}
\omega=\frac{4}{15a}\sqrt{\frac{2\epsilon}{15m}}=\frac{2}{w}\sqrt{\frac{2\epsilon}{15m}}
\eeq
where the width of the well is given by
\beq\label{2.20}
w=x_2-x_0=\frac{15a}{2}
\eeq
Therefore the decoherence time of equation (\ref{2.17}) can be expressed as
\beq\label{2.21}
\tau_{dec}=\frac{\hbar \sqrt{3}}{\sqrt{(E_i-E_f)(E_i-E_0)}}\left[\frac{2\hbar}{w K_B T}\sqrt{\frac{2\epsilon}{m}}\right]
\eeq
We will estimate the numerical value of decoherence time after deriving a certain relation between the decoherence time and Zeno time under some approximation.\\

We have already mentioned that, Quantum Zeno Effect is the slow down of quantum to classical transition due to frequent measurement. By definition this effect is something that slows down the process of decoherence. It is possible to control quantum to classical transition by frequent energy measurement. It has been shown \cite{10} that as a result of extremely frequent measurement, the system-reservoir coupling is eliminated and thus decoherence can be halted. Zeno time is the time scale within which the quantum states are frozen, ie the decay is halted. \\
\noindent In previous works \cite{8,11} we have derived the weak value of dwell time for interacting (via dissipation) systems, which is given by
\beq\label{3.1}
\begin{array}{ll}
\tau_w^D=\frac{\hbar}{\sqrt{(E_i-E_f)(E_i-E_0)}} \times\\
~~~~~~~\coth\left(\frac{\tau^M}{2\hbar}\sqrt{(E_i-E_f)(E_i-E_0)}\right)
\end{array}
\eeq
where $\tau^M$ is the measurement time. Now in explaining ``Hartman effect" \cite{12}, dwell time can be interpreted as the lifetime of the decaying state in the barrier region \cite{13,17,18}. If the interval between consecutive measurements ($\tau^{M}$) is significantly smaller than the Zeno time ($\tau^{Z}$), then the dynamics of the decay slows down or even asymptotically halted \cite{19}. Given the condition
\beq\label{3.2}
\tau^{M}\ll \tau^{Z}
\eeq
we see that the lifetime ($\tau^{L}$) of the decaying system, the measurement time ($\tau^{M}$) and the Zeno time obey the relation
\beq\label{3.3}
\tau^Z\approx \sqrt{\tau^L \tau^M}
\eeq
If we consider the interpretation of dwell time as the lifetime of the decaying state, then we can set $\tau^{L}=\tau_w^{D}$. Now under the assumption
\beq\label{3.4}
\tau^M\ll \frac{2\hbar}{\sqrt{(E_i-E_f)(E_i-E_0)}}
\eeq
using (\ref{3.1}) and (\ref{3.3}) we can get the Zeno time \cite{11} as
\beq\label{3.5}
\tau^Z=\frac{\sqrt{2}\hbar}{\sqrt{(E_i-E_f)(E_i-E_0)}}
\eeq
Referring to the equation (\ref{3.5}), we find that the approximation (\ref{3.4}) is nothing but
\beq\label{3.6}
\tau^{M}\ll \sqrt{2} \tau^{Z}
\eeq
which is similar to the approximation (\ref{3.2}). This condition can be fulfilled, if the initial and final energy states (with eigenvalues $E_i$ and $E_f$ respectively) of the decaying system are closely spaced. Now comparing equation (\ref{2.21}) and (\ref{3.5}) we get
\beq\label{3.7}
\frac{\tau_{dec}}{\tau^Z}= \frac{2\hbar}{w K_{B} T}\sqrt{\frac{3\epsilon}{m}}
\eeq
The preservation of quantum coherence leads us to conclude that Zeno time represents a certain lower limit of decoherence time, beyond which the system loses it's ``quantumness". The reason behind this statement lies in the definition of Zeno time itself. If the measurement is frequent enough that $\tau^M \ll \tau^Z$, we observe the asymptotic halting of state decay, hence preserves the coherence. But if the decoherence time is shorter than the Zeno time, the state will decay even within that estimated Zeno interval. As a result we will not be able to preserve the quantum coherence even by frequent measurement. From equation (\ref{3.7}) we find that imposition of this lower limit to the decoherence time leads us to a certain transitional temperature
\beq\label{3.8}
T_{tran}=\frac{2\hbar}{w K_{B} }\sqrt{\frac{3\epsilon}{m}}
\eeq
Above this temperature the coherence of the system can't be preserved even by Zeno dynamics. We have calculated this transitional temperature for our model system of trapped Be atom. Since the periodicity of the optical lattice is generally in the micrometer range and $\epsilon~(=h\nu_0)$ is about $8.5\times 10^{-25}$ Joules, we get that the transitional temperature is almost 200 microKelvin for Be atom. Hence above this temperature, decoherence will dominate the whole scenario. This is certainly an achievable temperature in case of Raman cooling. In case of Raman cooling for optical ion trap \cite{20}, the cooling temperature of Na atom is found to be around 1 microKelvin temperature, which is 0.42 times of the photon recoil temperature  $T_{rec}=\hbar^2\mathbf{k}^2/mK_{B}$. Since the recoil temperature is inversely proportional to the mass of the atom ($m$), for a lighter atom like Be, even lower cooling temperature can be achieved. It cannot be concluded that below the transitional temperature there will not be any loss of coherence. But at least we can predict from our calculation that decoherence should not dominate the scenario. As the decoherence time becomes larger than Zeno time, ``quantumness" of the system can be preserved within the Zeno interval. \\
\noindent Let us now calculate the decoherence time for a cooling temperature of 5 microKelvin. Consider the transition $|0\rangle|\uparrow\rangle \rightarrow |0\rangle|\downarrow\rangle$. In this case the Zeno time will be  $\tau^Z_w=\frac{\sqrt{2}\hbar}{\epsilon}$ \cite{11}. Subsequently, the decoherence time is found to be
\beq\label{3.9}
\tau_{dec}^{0^{+}\rightarrow 0^{-}}= \frac{2\hbar^2}{wK_B T}\sqrt{\frac{6}{m\epsilon}}
\eeq
Substituting the corresponding values of the parameters we get that the decoherence time is about 7 nanoseconds, whereas the Zeno time is found to be 0.17 nanoseconds. Similarly the timescales for other transitions can also be calculated.\\

\section{Conclusion}

Robust quantum memories are essential to realize the potential advances in quantum computation. Optical ion trap can be realized as a quantum storage device. But it is also essential to protect the information, which can be lost due to environmental \textbf{\emph{interaction}} in the form of decoherence. So it is very important to control the decohering effect in order to build an effective ion trap quantum computer. In this work, we have dealt with the question that whether and under what condition environment induced decoherence can be minimized. As we have discussed that in our understanding, the intrinsic relation between decoherence and Quantum Zeno effect can be exploited in this aspect. Frequent nonselective measurement forces the system to evolve in the reduced zeno subspaces, which can be considered as some ``quasi-isolated" system. If the Zeno effect (characterized by it's corresponding timescale ) is strong enough, so that the reduced subspaces remains quasi-isolated even under the influence of environmental interaction, effect of decoherence can be controlled. Based on this theoretical understanding, we have calculated a certain transitional temperature, by comparing the decoherence and Zeno timescales. It is clear from the above analysis that below this transitional temperature we can increase the decoherence time by controlling the parameters ($w,\epsilon$). Hence we can minimize the decohering effect asymptotically, though it can never be eliminated completely.

\end{document}